 \documentclass[preprint]{aastex631}
\usepackage{xspace} % intelligent space: space when there is word next, no-space when followed by a period

\newcommand{\Msun}{{\mathrm{M}_{\odot}}}

\newcommand{\Ha}{\mathrm{H}\alpha}
\newcommand{\Hb}{\mathrm{H}\beta}
\newcommand{\OI}{[\mathrm{O\,I}]}
\newcommand{\OII}{[\mathrm{O\,II}]}
\newcommand{\OIII}{[\mathrm{O\,III}]}
\newcommand{\NII}{[\mathrm{N\,II}]}
\newcommand{\SII}{[\mathrm{S\,II}]}
\newcommand{\HI}{\mathrm{H\,I}}
\newcommand{\HII}{\mathrm{H\,II}}

\begin{document}
%\title{O3 Equivalent Width as an Optical Diagnostic for Ionization Parameters in Metal Poor Galaxies}
\title{The $\OIII\lambda5007$ Equivalent Width as an Ionization-Parameter Diagnostic Across Cosmic Time}

\author[0000-0002-4752-128X]{Dian P. Triani}
\affiliation{Institute for Theory and Computation, Harvard-Smithsonian Center for Astrophysics \\
Cambridge, MA 02138, USA}
\affiliation{Center for Astrophysics \text{\textbar} Harvard \& Smithsonian, 60 Garden Street, Cambridge, MA 02138, USA}

\author[0000-0001-8152-3943]{Lisa J. Kewley}
\affiliation{Institute for Theory and Computation, Harvard-Smithsonian Center for Astrophysics \\
Cambridge, MA 02138, USA}
\affiliation{Center for Astrophysics \text{\textbar} Harvard \& Smithsonian, 60 Garden Street, Cambridge, MA 02138, USA}

%% Note that the \and command from previous versions of AASTeX is now
%% depreciated in this version as it is no longer necessary. AASTeX 
%% automatically takes care of all commas and "and"s between authors names.

%% AASTeX 6.31 has the new \collaboration and \nocollaboration commands to
%% provide the collaboration status of a group of authors. These commands 
%% can be used either before or after the list of corresponding authors. The
%% argument for \collaboration is the collaboration identifier. Authors are
%% encouraged to surround collaboration identifiers with ()s. The 
%% \nocollaboration command takes no argument and exists to indicate that
%% the nearby authors are not part of surrounding collaborations.

%% Mark off the abstract in the ``abstract'' environment. 
\begin{abstract}

Nebular emission lines and stellar continua probe the conditions of $\HII$ regions. The optical emission line $\OIII\lambda 5007$ is important because it is observable across a wide redshift range, enabling studies of interstellar medium (ISM) conditions across cosmic time. We investigate the equivalent width of $\OIII\lambda 5007$ (EWO3) as an ionization-parameter diagnostic, providing an alternative to the $\OIII \lambda 5007 / \OII \lambda 3727$ (O32) ratio, which requires broad wavelength coverage and dust corrections that may not be available at high redshift. We employ the MAPPINGS photoionization code with input spectra from \textsc{BPASS}, comparing single-star and binary-inclusive stellar populations. We find that binary-inclusive populations sustain harder ionizing spectra and strong $\OIII\lambda 5007$ emission to later ages than single-star populations. EWO3 is highly sensitive to the ionization parameter and exhibits systematic dependencies on metallicity, gas pressure, and stellar population age. While these dependencies broadly follow those of O32, EWO3 is less sensitive to gas pressure in the low-metallicity, high-ionization regime, making EWO3 well suited for studies of metal-poor and high-redshift galaxies. Comparing the models with observations, we find that local galaxies are reproduced by lower-pressure models, whereas intermediate- and high-redshift galaxies require higher gas pressures, indicating evolving ISM conditions across cosmic time. We present metallicity- and pressure-dependent calibrations of EWO3 as an ionization-parameter diagnostic. When validated against ionization parameters derived from O32, the $\log(P/k)=6$ calibration provides the best agreement for local galaxies, while the $\log(P/k)=7$ calibration best reproduces intermediate- and high-redshift galaxies.

%We present a new diagnostic for the ionisation parameter in star-forming galaxies, the equivalent width of OIII optical line. We derived our parameterization using the \textsc{MAPPINGS} photoionisation model with stellar input from \textsc{BPASS} stellar synthesis code. The new diagnostic can accurately convey the ionization state of the gas even at low metallicity. 

\end{abstract}

%% Keywords should appear after the \end{abstract} command. 
%% The AAS Journals now uses Unified Astronomy Thesaurus concepts:
%% https://astrothesaurus.org
%% You will be asked to selected these concepts during the submission process
%% but this old "keyword" functionality is maintained in case authors want
%% to include these concepts in their preprints.

%% From the front matter, we move on to the body of the paper.
%% Sections are demarcated by \section and \subsection, respectively.
%% Observe the use of the LaTeX \label
%% command after the \subsection to give a symbolic KEY to the
%% subsection for cross-referencing in a \ref command.
%% You can use LaTeX's \ref and \label commands to keep track of
%% cross-references to sections, equations, tables, and figures.
%% That way, if you change the order of any elements, LaTeX will
%% automatically renumber them.
%%
%% We recommend that authors also use the natbib \citep
%% and \citet commands to identify citations.  The citations are
%% tied to the reference list via symbolic KEYs. The KEY corresponds
%% to the KEY in the \bibitem in the reference list below. 

\section{Introduction} \label{sec:intro}

In star-forming galaxies, young stars are enshrouded by their thick and dense birth clouds. These hot stars produce high-energy photons that ionize the surrounding gas, which consists mainly of hydrogen, creating the $\HII$ region. The photoionized and photoheated gas in the $\mathrm{H_{II}}$ region cools down via (i) recombination of free electrons and ions, and (ii) decay of collisionally excited electrons in atoms and ions, producing nebular emission line and continuum. These nebular components add to the composite spectrum of galaxies \citep[see][for a review]{Kewley19}. 

The observed properties of nebular emission depend on several parameters: (i) the production rate of ionizing Lyman continuum photons, which traces the current star formation rate; (ii) the density and pressure of the birth clouds; and (iii) the gas-phase metallicity. The first two parameters are encapsulated in the ionization parameter, $Q$, defined as the ratio of the ionizing photon flux to the hydrogen gas density. This ionization parameter, with units of $\mathrm{cm , s^{-1}}$, represents the maximum velocity of the ionization front driven by the radiation field. It connects to the dimensionless ionization parameter, $U \equiv Q/c$, where $c$ is the speed of light. 

Oxygen produces strong optical emission lines commonly used to diagnose various interstellar medium (ISM) properties. For example, the ratio of oxygen lines in different ionization states, $\OIII\lambda5007/\OII\lambda3727$ (hereafter O32), serves as a widely used proxy for the ionization parameter \citep[e.g.,][]{Alloin78, Baldwin81, Kewley02}. Combined with hydrogen, neon, and nitrogen emission lines, oxygen lines are widely used to infer the gas-phase metallicity of the ISM using both empirical \citep{Pagel79, Maiolino08, Pilyugin16} and theoretical calibrations \citep{Kewley02}. A commonly used metallicity diagnostic is the $R_{23}$ index, defined as $(\OIII\lambda\lambda4959,5007 + \OII\lambda3727)/\Hb$. Techniques based on oxygen emission lines also distinguish the dominant ionizing source in galaxies. For example, the BPT diagram \citep{Baldwin81} employs $\OIII\lambda5007/\Hb$ versus $\NII\lambda6584/\Ha$ to separate star-forming galaxies from those with significant AGN contributions \citep{Veilleux87, Kewley01, Kauffmann03}. Another commonly used diagnostic employs O32 versus $\OI\lambda6300/\Ha$ to identify ionization by fast radiative shocks.

The $\OIII\lambda5007$ line (hereafter O3) is especially important for observations of high-redshift galaxies during the epoch of reionization because of its strength. Strong O3 lines have been detected up to $z\approx9$ during the JWST commissioning \citep{Rigby22, Sun22}. While extreme emission line galaxies are rare in the local Universe (only $0.23\%$ of SDSS galaxies; \citealt{debarros19}), such populations are more common at high redshifts. The equivalent width (EW) of the O3 line (EWO3) increases with redshift. At $z\approx2$, typical EWs are $100$–$200 , \mathrm{\AA}$ \citep[e.g.,][]{Khostovan+16, Reddy18}, whereas at $z\approx6$–$8$, they reach $\sim700 , \mathrm{\AA}$ \citep{Labbe13, debarros19, Endsley21}. Individual galaxies have shown extreme nebular emission with EWs of $1000$–$2000 , \mathrm{\AA}$ \citep{Roberts-Borsani16, Tang19}. The prominence of strong line emitters at the epoch of reionization (EoR) suggests these galaxies contribute significantly to the ionizing photons responsible for reionizing the Universe.

The second ionization of oxygen, which produces the O3 line, requires ionizing radiation typically generated by hot, massive, young stars. However, our interpretation of such nebular spectra depends on the assumptions of the shape of the intrinsic spectra, which change with the stellar population model used. For example, binary evolution in the \texttt{BPASS} stellar population model \citep{Eldridge17} allows the second ionization of oxygen to occur in older stellar populations than previously expected \citep{Xiao18}. Similarly, stellar population models that include stellar rotation predict harder ionizing spectra and enhanced ionizing photon production persisting to later ages, extending to 4 Myr, compared to non-rotating models \citep{Byler17}.

The equivalent width of the O3 line provides insight into the stellar population and ISM gas properties. Empirical studies have found a positive correlation between EWO3 and the ionization parameter, as traced by the O32 ($\OIII\lambda5007/\OII\lambda3727$) ratio \citep{Nakajima14, Reddy18, Tang19}. A high ionization parameter is typically associated with low gas-phase metallicity, as metal-poor galaxies produce harder ionizing photons \citep{Schaerer03}. This correlation arises from the effects of metallicity on line blanketing and wind opacities in young, massive stars \citep{Kewley02, Pilyugin05, Dopita06}. A negative correlation between EWO3 and stellar mass is observed in both low- and high-redshift galaxies \citep{Khostovan+16, Reddy18}, potentially driven by the mass-metallicity relation \citep{Tremonti04}, where metallicity increases with stellar mass. Thus, galaxies with high EWO3 are typically characterized by high ionization parameters, young stellar ages, low metallicity, and low stellar mass.

With the increasing importance of strong line emitters at high redshifts and the launch of JWST targeting early-Universe galaxies, understanding this population is critical. Unfortunately, such systems are rarely captured in local spectroscopic surveys like SDSS \citep{Tremonti04}, which are biased toward higher stellar mass and luminosity. However, efforts have been made to study local metal-poor systems with extreme emission line properties \citep[e.g.,][]{Nakajima22}.

In this study, we explore the equivalent width of the O3 line and its dependence on $\mathrm{H_{II}}$ region properties using theoretical photoionization models. This work complements empirical studies of low- and high-redshift galaxies \citep{Khostovan+16, Reddy18, Tang19}. Section~\ref{sec:sample} describes the comparison samples. Section~\ref{sec:modeling} presents the photoionization models, while Section~\ref{sec:spectra} examines the ionizing radiation field. Section~\ref{sec:EW} investigates the dependence of EWO3 on the stellar population and ISM properties, and Section~\ref{sec:diagnostic} explores the EWO3--O32 diagnostic space through comparisons with observations. Section~\ref{sec:calibration} presents the EWO3 ionization-parameter calibration. We summarize our conclusions in Section~\ref{sec:summary}.

\section{Comparison Sample}\label{sec:sample}
The local galaxy sample is drawn from the \textit{Sloan Digital Sky Survey Data Release 8} \citep[SDSS DR8;][]{Aihara11}. We select star-forming galaxies in the redshift range $0.04 < z < 0.1$ to minimize aperture effects \citep{Kewley05}, requiring a signal-to-noise ratio greater than 5 in $\Hb$, $\OIII\lambda5007$, $\Ha$, and $\NII\lambda6584$. To exclude contamination from active galactic nuclei (AGN), we follow the classification scheme of \citet{Kewley06}, which combines the \citet{Kauffmann03} demarcation in the $\NII\lambda6584/\Ha$ versus $\OIII\lambda5007/\Hb$ diagram with the \citet{Kewley01} demarcations in the $\SII\lambda\lambda6717,31/\Ha$ versus $\OIII\lambda5007/\Hb$ and $\OI\lambda6300/\Ha$ versus $\OIII\lambda5007/\Hb$ diagrams.

Because the O32 ratio spans a wide wavelength baseline, we correct nebular emission-line fluxes for dust attenuation using the Balmer decrement. We assume an intrinsic $\Ha$/$\Hb$ ratio of 2.86, appropriate for Case B recombination at a temperature of $10^4$~K and an electron density of $100~\mathrm{cm}^{-3}$, and attribute deviations from this value to dust attenuation. We compute the color excess, $E(B-V)$, on a galaxy-by-galaxy basis using the Milky Way extinction curve of \citet{Fitzpatrick19}, implemented with the \texttt{dust\_extinction} Python package \citep{2013A&A...558A..33A}. Dust-corrected emission-line fluxes are then used to compute the O32 ratio. After applying all selection criteria and dust corrections, the final SDSS comparison sample contains 89{,}308 galaxies.

We compare our model predictions with several compilations of intermediate- and high-redshift observations from \citet{Reddy18}, \citet{Tang19}, \citet{Fujimoto23}, and \citet{Tang23}. The sample from \citet{Reddy18} is drawn from the \textit{MOSFIRE Deep Evolution Field} survey \citep[MOSDEF;][]{Kriek15} and consists of 1134 star-forming galaxies spanning $1.4 \lesssim z \lesssim 3.8$. From this sample, 214 galaxies have reliable EWO3 and O32 measurements and are used for comparison with the models.

From the compilation of extreme line emitters presented in \citet{Tang19}, 54 galaxies at $1.3 < z < 2.4$ have reliable EWO3 and O32 measurements and are used for comparison with the models. We also compare our model predictions to galaxies at $z \gtrsim 7.5$ observed with \textit{JWST}/NIRCam and \textit{JWST}/NIRSpec as part of the Cosmic Evolution Early Release Science Survey (CEERS; ERS 1345, PI: S. Finkelstein) from \citet{Fujimoto23}, for which six galaxies have O32 constraints. Finally, we include 11 galaxies at $z \gtrsim 8$ from \citet{Tang23}, observed with \textit{JWST}/NIRSpec in the CEERS field. For these systems, only blended $\mathrm{EW}(\OIII+\Hb)$ measurements are available; we reconstruct EWO3 from the blend using the measured $\OIII$/H$\beta$ flux ratios (Section~\ref{sssec:nonlocal}) and use these galaxies as a qualitative high-redshift consistency check on the EWO3 ionization-parameter calibration.

In addition, we include extremely metal-poor galaxies (EMPGs) from the EMPRESS survey \citep{Nakajima22}. EMPRESS combines newly discovered EMPGs from Subaru/HSC imaging and follow-up spectroscopy with metal-poor galaxies from the literature. Applying quality cuts requiring reliable EWO3 and O32 measurements, we retain 103 EMPRESS galaxies for comparison with the models. These galaxies lie predominantly at $z < 0.05$ and span $6.9 \lesssim 12+\log(\mathrm{O/H}) \lesssim 8.9$, with metallicities measured using the direct $T_{\mathrm e}$ method. For consistency in the ionization-parameter analysis, we recompute gas-phase metallicities using the R23-based calibration of \citet{Kobulnicky04} (hereafter KK04) for all samples except \citet{Tang19}, for which individual gas-phase metallicities are unavailable. Consequently, the \citet{Tang19} sample is used only for comparison in the EWO3--O32 diagnostic plane.

\section{Modelling} \label{sec:modeling}
The observed emission lines arise from the interaction between ionizing photons and nebular gas. The number of ionizing photons radiated by the stars depends on the stellar population models, which we describe in Section \ref{ssec:sps}. The stellar spectra are used as input for the photoionization code \textsc{MAPPINGS V}\citep{Sutherland18}, along with other parameters that describe the gas conditions. We describe our setup for the \textsc{MAPPINGS V} model in Section \ref{ssec:mappings}. 

\subsection{Stellar population synthesis} \label{ssec:sps}
We adopt the ionizing stellar radiation field constructed using the \textit{Binary Population and Spectral Synthesis} (\textsc{BPASS}) v2.2 code \citep{Stanway18}. \textsc{BPASS} generates the spectral energy distribution (SED) of a coeval stellar population, in which all stars share the same age and metallicity and are assumed to have formed from a single molecular cloud. For a given age and metallicity, the code populates stellar evolutionary tracks using a stellar mass distribution determined by a chosen initial mass function (IMF), which describes the number density of stars per unit mass at the zero-age main sequence. From the IMF, stars of different masses evolve off the main sequence at varying timescales, following distinct evolutionary tracks based on their mass. Extensive libraries of stellar spectra \citep[e.g.,][]{Lejeune98, Westera02} are used to assign spectra to each star based on its computed evolutionary properties. The overall spectrum of the coeval stellar population is obtained by integrating the individual spectra. Spectra of multiple populations are then combined, based on an assumed star formation history (SFH), to produce the composite SED of a galaxy.

Unlike traditional stellar population synthesis codes that only consider single stellar evolution, \textsc{BPASS} incorporates binary evolution models alongside the latest stellar atmosphere models. In modeling binary evolution, \textsc{BPASS} accounts for mass transfer between binary stars, requiring input parameters such as the IMF, the initial binary mass ratio distribution, and the initial period ratio distribution \citep{Eldridge17, Stanway18}. Binary populations are particularly important for modeling metal-poor galaxies with high equivalent widths, as binaries increase the ionizing flux by $50\text{–}60\%$ at low metallicities ($0.05 Z_\odot \leq Z \leq 0.3 Z_\odot$), compared to an $10\text{–}20\%$ increase at higher metallicities. At older ages, stellar populations that include binaries exhibit bluer spectra than those consisting of single stars alone. 

\textsc{BPASS} uses single and binary stellar evolution tracks from the \textsc{STARS} code \citep{Eggleton71, Eldridge08}. The code follows the evolution of single and binary stars from the main sequence phase to the end of their lives. Synthetic stellar spectra are derived from pre-calculated grids of stellar atmospheres matched to the stellar evolution models. \textsc{BPASS} uses the BaSel v3.1 stellar atmosphere library \citep{Westera02} for stars with metallicities in the range $\mathrm{[Fe/H]} = -2.0$ to $0.5$, supplemented by the earlier version of \citet{Lejeune98} to extend coverage to lower metallicities down to $\mathrm{[Fe/H]} = -3.5$. For stars with surface hydrogen fractions below 0.4 and surface temperatures above $\log(T/\mathrm{K}) = 4.45$, \textsc{BPASS} categorizes them as Wolf-Rayet stars and uses spectra from the Potsdam PoWR theoretical atmosphere library \citep{Hamann06, Sander15}. Spectra for O stars are calculated using the WM Basic code \citep{Pauldrach01}, based on the grid created by \citet{Smith02}.

\textsc{BPASS} has nine IMF options with slope and stellar mass range variations. In this work, we use the IMF described as a double power law with a slope of $\alpha_1 = -1.30$ for stellar mass ranging from $0.1$ to $0.5 \Msun$ and a slope of $\alpha_2 = -2.35$ for stellar mass ranging from $0.5 \Msun$ to $100 \Msun$. This option closely resembles the \citet{2001Kroupa} IMF which consists of three power laws with slopes $\alpha_1 = 0.3$, $\alpha_2 = -1.3$ and $\alpha_3 = -2.3$ for stellar mass ranges of $0.01 - 0.08 \Msun$, $0.08 - 0.5 \Msun$, and $0.5 - 120 \Msun$, respectively.

When assembling spectra from \textsc{BPASS} to construct a galaxy SED, an assumed star formation history is required. The SFH of a galaxy is typically stochastic, with multiple star formation episodes spanning various timescales. However, nebular emission line models are especially sensitive to the ionizing emission from short-lived massive stars, allowing simplified models to approximate the dominant contributions. In this work, we adopt an instantaneous burst SFH, consisting of $10^6\ \Msun$ of stars formed at a single epoch. The properties of the emission lines vary with the age of the stellar population following the burst until the stellar population ceases producing ionizing photons. The age at which ionizing photon production stops differs significantly between models assuming only single stars and those including binary contributions \citep{Xiao18, D'agostino19}. Models incorporating binaries sustain ionizing photon production to significantly older ages than those assuming single stars alone \citep{Xiao18}.

In this work, we explore how variations in the input stellar spectra affect the resulting equivalent width of O3. The stellar population parameters explored are as follows:

\textbf{Stellar population model:} We investigate the influence of binary evolution by comparing results obtained using the \textsc{BPASS} v2.2 models that include binary stellar populations with those computed using the single-star-only variant of \textsc{BPASS} v2.2.

\textbf{Age:} We adopted instantaneous burst models of $10^6\ \Msun$ stars with ages of $\log(\mathrm{Age/Myr}) = 0, 0.5, 1, 1.5, 2$.

\textbf{Stellar metallicity:} We use five different metallicities, $Z = 0.001, 0.004, 0.008, 0.02$ and $0.04$.

\subsection{Photoionization model of H II region} \label{ssec:mappings}
We use the \textsc{MAPPINGS V}\footnote{\url{https://mappings.anu.edu.au/code/}} photoionization code \citep{Sutherland18, Nicholls14}, which models the heating, ionization, and shocks within a gas nebula. \textsc{MAPPINGS V} calculates the ionization fractions of elements at various radii within the nebula \citep{Sutherland93}, incorporating a self-consistent dust treatment \citep{Groves04}. The latest version of the code includes updated atomic data from the CHIANTI Version 8 database \citep{DelZanna15}, incorporates elemental depletion factors from \cite{Jenkins14}, and improves the integration functions in ionization rate calculations to enhance accuracy.

\textsc{MAPPINGS V} requires ionizing spectra and the physical conditions of the nebular gas as inputs. Changes in stellar spectra and gas conditions affect the resulting nebular spectra, including both the continuum and emission line flux. Below, we describe how we vary the properties of the nebular gas:

\textbf{Ionization parameter:} We adopt ionization parameters ($Q / \mathrm{cm}\ \mathrm{s^{-1}}$) in a logarithmic range from $6.5$ to $9.0$, with increments of $0.5$ dex. This corresponds to the dimensionless paramater of $\log U = -4$ to $-1.5$ with the same increments. These values are consistent with observed ionization parameters in local starburst galaxies \citep{Rigby04, Kewley01}

\textbf{Gas metallicities:} We consider a range of gas metallicities that match the stellar population models: $Z = 0.001, 0.004, 0.008, 0.02$ and $0.04$. These are identified as z001, z004, z008, z02 and z04, respectively. We use the solar abundance from \citet{Anders89}, which adopts a solar metallicities of $Z = 0.02$ and $12 + \log (\mathrm{O/H}) = 8.93$. Table \ref{tab:abundance} lists the elemental abundances and depletion factors. Abundances for other metallicities are scaled linearly, except for helium, which is held constant. The depletion factors for each element are from \citet{Jenkins14}, using a logarithmic base depletion of 1.50 for iron. These depletion factors represent fixed relative depletion patterns and match those observed in the diffuse ISM of the Milky Way; the same depletion fractions are applied at all metallicities, such that the absolute amount of depleted material scales with metallicity.

\textbf{Dust treatment:} \textsc{MAPPINGS V} treats dust self-consistently, including the extinction and scattering of stellar radiation by dust grains, photoelectric heating of the gas by dust, grain charging, and other dust-related processes \citep{Groves04}. In this work, we adopt the grain size distribution of \citet{Mathis77}, disable grain destruction, and set a carbon depletion fraction in polycyclic aromatic hydrocarbon molecules (PAHs) to $0.3$.

\textbf{Pressure and density:} The ideal gas law links pressure, density, and temperature in an ideal gas system. Here, we adopt an isobaric system, meaning the pressure remains constant throughout the gas cloud. As the gas temperature drops near the nebula's edge, due to its distance from the radiation source, the gas density increases, affecting the ion density and nebular emission. We adopt an isobaric equation of state and explore constant gas pressures of $\log(P/k)=5$, 6, 7, and $8\ \mathrm{cm^{-3}\ K}$. For each model, the initial gas temperature is set to $T=10^4\ \mathrm{K}$, corresponding to initial hydrogen densities of 10, 100, $10^3$, and $10^4\ \mathrm{cm^{-3}}$, respectively. The lowest-pressure model is broadly consistent with the characteristic electron density of $26.8^{+0.2}_{-0.2}\ \mathrm{cm^{-3}}$ measured for local galaxies in the COSMOS-[O\,\textsc{ii}] sample \citep{Kaasinen17}. We additionally explore higher gas pressures to investigate conditions representative of more extreme star-forming environments, including high-redshift galaxies.

\textbf{Geometry and boundedness:} We assume a closed, spherical, radiation-bounded nebula. In this configuration, the radiation source is enclosed within the nebula, and all ionizing photons are absorbed by the gas. The calculation terminates once $99\%$ of the hydrogen has recombined. This assumption is appropriate for typical $\HII$ regions in galaxies with small ionizing-photon escape fractions. In contrast, an open geometry allows radiation to escape from the modeled nebula, while a density-bounded model represents a nebula that is truncated before all ionizing photons are absorbed, as may occur in leaky $\HII$ regions \citep[e.g.,][]{Zastrow13}. A plane-parallel geometry is instead appropriate when the ionizing source is external to the modeled gas cloud, as commonly assumed for the narrow-line regions of AGNs.

The stellar and nebular parameters explored in the photoionization model grid are summarized in Table~\ref{tab:model_grid}.

\begin{table}
\centering
\caption{Summary of stellar and nebular parameters explored in the photoionization model grid.}
\label{tab:model_grid}
\begin{tabular}{lll}
\hline
Parameter & Symbol & Values explored \\
\hline
Stellar evolution model & SSP & Binary, single-star (BPASS v2.2) \\
Star formation history & SFH & Instantaneous burst ($10^6~\Msun$) \\
Stellar age & Age & $\log(\mathrm{Age/Myr}) = 0, 0.5, 1, 1.5, 2$ \\
Stellar metallicity & $Z_\ast$ & 0.001, 0.004, 0.008, 0.02, 0.04 \\
Gas metallicity & $Z$ & 0.001, 0.004, 0.008, 0.02, 0.04 \\
Ionization parameter & $\log U$ & $-4$ to $-1.5$ (step 0.5) \\
Gas pressure & $\log(P/k)$ & 5, 6, 7, 8 ($\mathrm{cm^{-3}\,K}$) \\
Geometry & -- & Closed spherical, radiation-bounded \\
Dust model & -- & MRN grain size distribution \\
\hline
\end{tabular}
\end{table}

\begin{table}
    \centering
    \begin{tabular}{c|c|c}
    %\hline
         Element & $\log(\mathrm{Z_\odot})$ & $\log(\mathrm{D})$\\
         \hline
         \hline
         H & 0.00 & 0.00 \\
         He & -1.01 & 0.00 \\
         Li & -10.84 & -0.22 \\
         Be & -10.85 & -0.40 \\
         B & -9.40 & -0.58 \\
         C & -3.44 & -0.16 \\
         N & -3.95 & -0.04 \\
         O & -3.07 & -0.11 \\
         F & -7.44 & -0.09 \\
         Ne & -3.91 & 0.00 \\
         Na & -5.67 & -0.42 \\
         Mg & -4.42 & -0.70 \\
         Al & -5.53 & -0.70 \\
         Si & -4.45 & -0.71 \\
         P & -6.55 & -0.11 \\
         S & -4.79 & 0.00 \\
         Cl & -6.73 & -0.09 \\
         Ar & -5.44 & 0.00 \\
         K & -6.88 & -0.62 \\
         Ca & -5.64 & -1.95 \\
         Sc & -8.90 & -0.69 \\
         Ti & -7.01 & -1.95 \\
         V & -8.00 & -2.17 \\
         Cr & -6.32 & -1.45 \\
         Mn & -6.61 & -1.27 \\
         Fe & -4.33 & -1.50 \\
         Co & -7.07 & -1.64 \\
         Ni & -5.75 & -1.57 \\
         Cu & -7.79 & -0.90 \\
         Zn & -7.40 & -0.20
    \end{tabular}
    \caption{Adopted solar abundance set from \citet{Anders89} and depletion factors for each element. The depletion values are computed following \citet{Jenkins14} using a logarithmic base of iron depletion $= 1.50$}
    \label{tab:abundance}
\end{table}

\section{Ionizing radiation field}
\label{sec:spectra}

\begin{figure}
    \centering
    \includegraphics[width=0.9\linewidth]{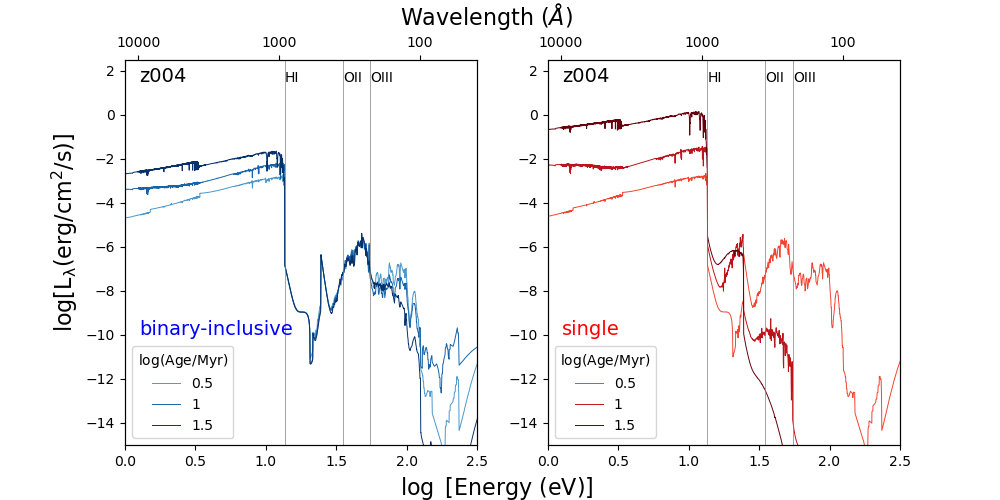}
    \caption{Ionizing spectra from \textsc{BPASS} v2.2 stellar population models at a fixed metallicity of $Z=0.004$. The left panel shows spectra from binary-inclusive models, while the right panel shows spectra from single-star models. For clarity, only three representative stellar population ages, $\log(\mathrm{Age/Myr})=0.5$, 1, and 1.5, are shown, although the full model grid spans $\log(\mathrm{Age/Myr})=0$--2. Vertical gray lines mark the ionization potentials of $\HI$, $\OII$, and $\OIII$.}
    \label{fig:bpass_age}
\end{figure}

Stellar populations produce high-energy photons which can ionize the gas in $\mathrm{H_{II}}$ regions of galaxies. The parameters of the stellar population model, such as age, metallicity, and inclusion of binary stars, govern the number and the spectral slope of the ionizing photons, which are blueward of the $912\ \mathrm{\AA}$ Lyman limit. In this section, we discuss the effects of changing stellar population parameters on the shape of the SED.

\subsection{Time evolution}

The effective temperature of stars controls the energy distribution of the photons they emit. Young, massive stars produce large numbers of high-energy ultraviolet photons capable of ionizing hydrogen and oxygen, whereas older and lower-mass stars emit predominantly at longer wavelengths. As a result, the ionization state of nebular gas in star-forming galaxies depends sensitively on the time evolution of the stellar population illuminating the gas.

Figure~\ref{fig:bpass_age} shows the time evolution of the ionizing spectra predicted by \textsc{BPASS} v2.2 at a fixed metallicity of $Z=0.004$, comparing binary-inclusive models (left panel) with single-star models (right panel). For clarity, only three representative stellar population ages, $\log(\mathrm{Age/Myr})=0.5$, 1, and 1.5, are shown, although the full analysis includes models spanning $\log(\mathrm{Age/Myr})=0$--2. Vertical gray lines indicate the ionization potentials of $\HI$, $\OII$, and $\OIII$.

In single-star models, the ionizing radiation field is dominated at early times by massive O stars. The main-sequence lifetime of O stars is approximately $2.5$–$4$ Myr \citep{Lamers91}, after which ionization can briefly persist as some massive stars evolve into Wolf–Rayet (WR) stars through strong stellar winds that strip their hydrogen-rich envelopes. WR stars are hotter than O stars but short-lived, providing ionizing photons for only $\sim0.5$ Myr \citep{Lamers91}. Consequently, single-star stellar populations can sustain significant ionization for only $\sim3$–$5$ Myr, a period often referred to as the equilibrium age \citep{Kewley01}. At later times, the ionizing flux declines rapidly as O stars leave the main sequence and cooler B stars come to dominate the spectrum. By ages of $\sim6$ Myr, the production of photons capable of ionizing $\OII$ and $\OIII$ becomes negligible in instantaneous-burst populations, such as individual stellar clusters or $\HII$ regions \citep{Wofford16}.

In contrast, binary-inclusive stellar populations exhibit a markedly different time evolution. As shown in Figure~\ref{fig:bpass_age}, binary-inclusive models produce ionizing spectra similar to single-star models at young ages but retain substantially harder radiation fields at later times. This extended production of high-energy photons arises from binary evolution processes such as mass transfer through Roche-lobe overflow, stellar mergers, and envelope stripping during the red supergiant phase \citep{Dray07, Schneider16, Smith14}. These processes can rejuvenate stars or expose hot helium cores, leading to the formation of WR or helium stars at ages when single-star populations have already ceased producing hard ionizing radiation. As a result, binary-inclusive models sustain photons capable of ionizing $\mathrm{O\,II}$ and $\mathrm{O\,III}$ to significantly later ages, in some cases up to tens of Myr \citep{Xiao18, D'agostino19}.

The prolonged hardness of the ionizing radiation field in binary-inclusive models has direct consequences for nebular emission, enabling strong $\mathrm{[O\,III]}$ emission and large EWO3 values to persist in older stellar populations. This behavior underpins the strong dependence of EWO3 on stellar population age and motivates the use of binary-inclusive stellar population models in the analysis presented in the following sections.

\begin{figure}
    \centering
    \includegraphics[width=0.9\linewidth]{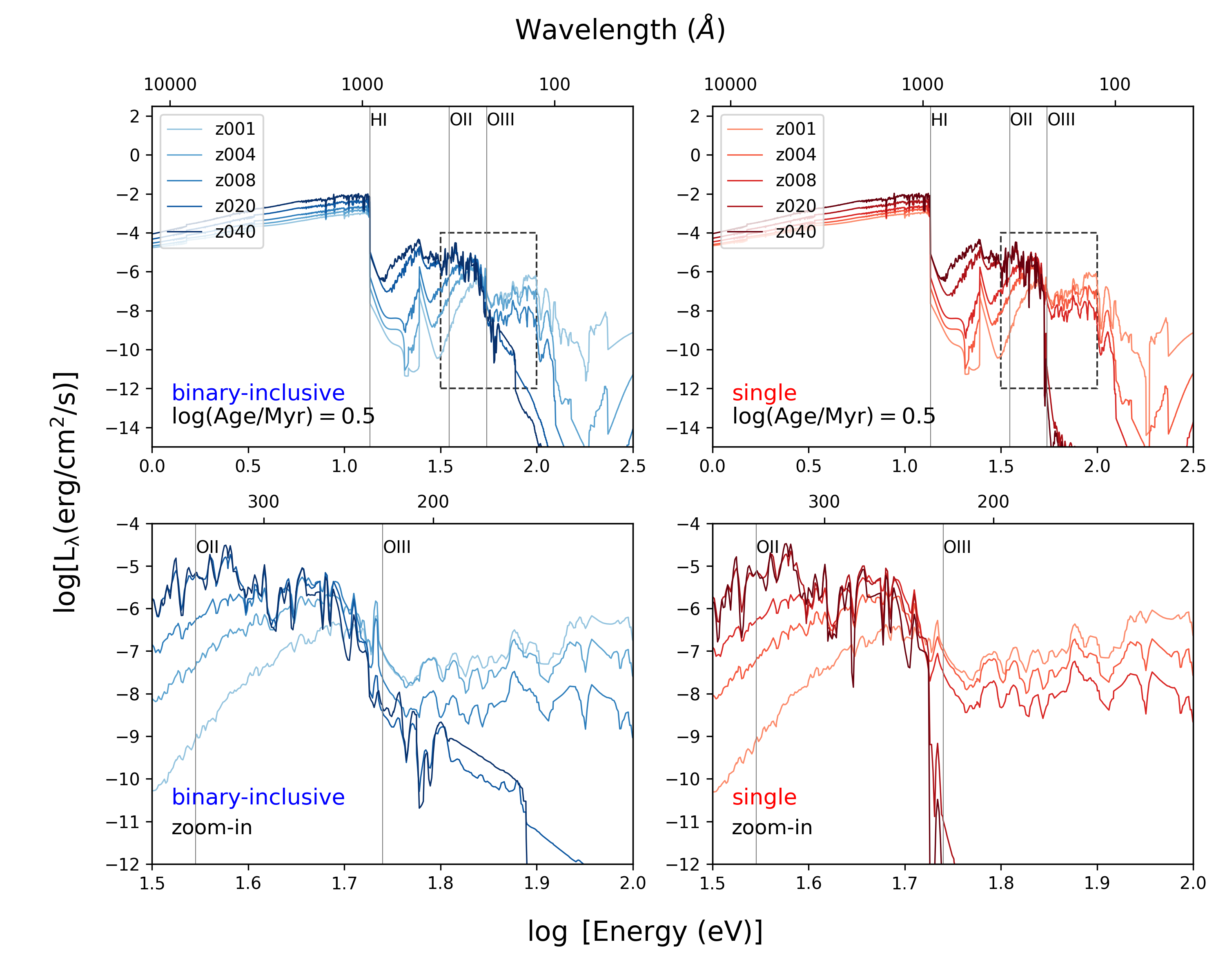}
    \caption{Ionizing spectra from \textsc{BPASS} v2.2 stellar population models at a fixed age of $\log(\mathrm{Age/Myr})=0.5$ (3.2 Myr). The left panels show binary-inclusive models, while the right panels show single-star models. Different line colors indicate stellar metallicities ranging from $Z=0.001$ to $Z=0.04$. Vertical gray lines mark the ionization potentials of $\HI$, $\OII$, and $\OIII$. The lower panels show a zoom-in of the region highlighted by the dashed box in the upper panels, illustrating the differences in the ionizing continuum around the $\OII$ and $\OIII$ ionization thresholds. Binary-inclusive models retain a harder ionizing spectrum beyond the $\OIII$ ionization threshold than single-star models.}
    \label{fig:bpass_Z}
\end{figure}

\subsection{Metallicity evolution} \label{ssec:metal}

At fixed stellar age, increasing metallicity leads to a systematic decrease in the number of ionizing photons in both binary-inclusive and single-star stellar population models, as shown in Figure~\ref{fig:bpass_Z}. This trend arises from two primary effects. First, metals in stellar atmospheres absorb stellar radiation, including hard ionizing photons. A higher metal abundance increases atmospheric opacity, lowering stellar surface temperatures and reducing the production rate of ionizing photons. Second, metallicity influences stellar evolution. Metal-poor stars experience weaker stellar winds and reduced mass loss, resulting in longer main-sequence lifetimes compared to metal-rich stars. These extended lifetimes allow metal-poor populations to continue producing ionizing photons at later times \citep{Eldridge11, Eldridge12, Xiao18}. As a result, at an age of 3.2 Myr, low-metallicity stellar populations produce substantially harder and more intense ionizing spectra than their high-metallicity counterparts.

Although the ionizing flux decreases with increasing metallicity in both stellar population models, the effect is less pronounced in the binary-inclusive case. Binary interactions, including mass transfer and envelope stripping, can partially counteract the effects of stellar winds and metallicity on stellar evolution and opacity. Consequently, at higher metallicities ($Z = 0.02$ and $0.04$), binary-inclusive models continue to produce photons with energies exceeding the ionization potential of $\OIII$, whereas single-star models show a much sharper decline in hard ionizing radiation. The zoomed panels of Figure~\ref{fig:bpass_Z} highlight this behavior, showing that binary-inclusive populations retain a substantially harder ionizing continuum than their single-star counterparts beyond the $\OIII$ ionization threshold.

\section{Equivalent width and ionization parameter}\label{sec:EW}
\label{sec:EWO3-Q}
The observed emission lines arise from the interaction between ionizing radiation and nebular gas. Traditionally, nebular diagnostics rely on emission-line ratios to infer physical properties such as the ionization parameter, gas-phase metallicity, and electron temperature \citep[see][for a review]{Kewley19}. In this work, however, we investigate the equivalent widths of strong emission lines as alternative diagnostics. Because EWs compare the emission-line flux to the adjacent stellar continuum, they do not require absolute flux calibration and are less sensitive to dust attenuation over the narrow wavelength interval surrounding the line. They are particularly useful when only a limited spectral range is available, as is often the case for high-redshift galaxies.

Several empirical studies have reported correlations between O32 and the equivalent widths of strong nebular emission lines across a wide range of redshifts \citep{Reddy18,Tang19,Tang23}. These correlations suggest that equivalent widths may serve as alternative tracers of the ionization parameter. However, it is not obvious that the equivalent width of O3 is uniquely suited for this purpose. Other strong optical emission lines, including $\OII$, $\Hb$, and $\Ha$, may also respond to changes in the ionizing radiation field. We therefore compare the equivalent widths of these lines using our photoionization models to determine which provides the most robust ionization-parameter diagnostic. As a benchmark, we compare the equivalent-width diagnostics with the widely used O32 ratio, which traces the ionization parameter through the relative abundance of singly and doubly ionized oxygen \citep[e.g.,][]{Alloin78,Baldwin81,Kewley02}.

\begin{figure}
    \centering
    \includegraphics[width=1.0\linewidth]{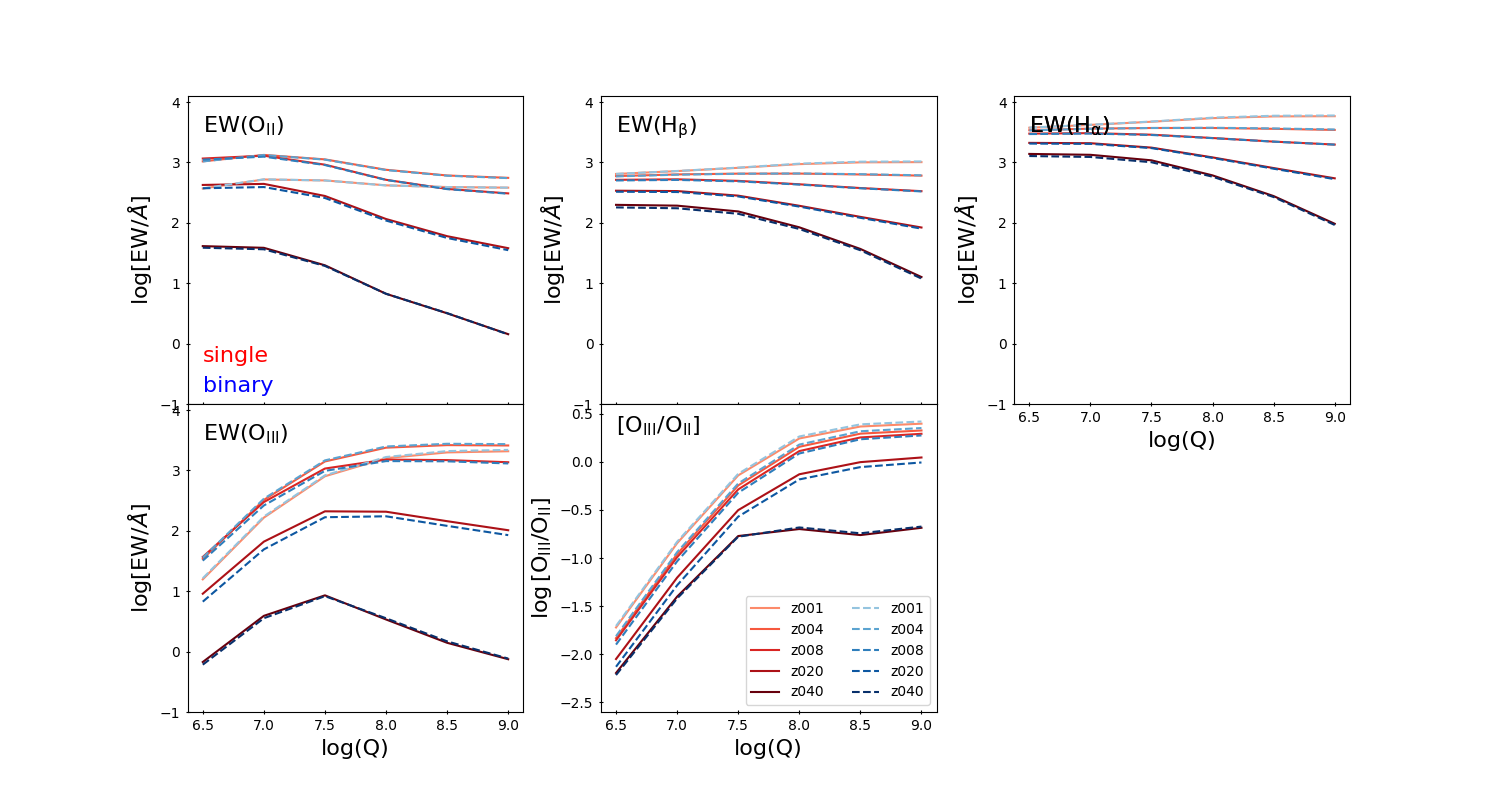}
    \caption{Equivalent widths of $\OII$, $\Hb$, $\Ha$, O3, and the O32 ratio ($\OIII\lambda5007/\OII\lambda3727$) as a function of ionization parameter, $\log U$. The curves are derived from models assuming $\log(\mathrm{Age/Myr})=0.5$. Red solid lines show the single-star models, while blue dashed lines show the binary-inclusive models. Different colors indicate gas metallicity, as labeled in the legend.}
    \label{fig:EW_Q}
\end{figure}

Figure~\ref{fig:EW_Q} compares the response of the equivalent widths of $\OII$, H$\beta$, H$\alpha$, and O3 to changes in ionization parameter. As a benchmark, the final panel shows the O32 ratio, the widely used ionization-parameter diagnostic. Results are shown for both single-star (red solid lines) and binary-inclusive (blue dashed lines) stellar population models, with different line shades indicating stellar metallicity as labeled in the legend. 

Across all metallicities, O32 increases steeply with increasing $\log(Q)$ up to $\log(Q)\simeq7.5$, above which the relation flattens and becomes non-monotonic at super-solar metallicities. This behaviour is consistent with the classical dependence of O32 on ionization parameter in photoionization models, where enhanced cooling and changes in the ionization balance become increasingly important at high metallicity and high $Q$ \citep[e.g.,][]{Kewley02}. At the fixed age considered here, the O32--$Q$ relation shows only minor differences between the single-star and binary-inclusive models.

In contrast to O32, the equivalent widths of O2, H$\beta$, and H$\alpha$ exhibit weak or non-monotonic trends with $\log(Q)$ over most of the explored parameter space. In particular, these equivalent widths remain approximately constant and only decline at high $\log(Q)$ for near-solar and super-solar metallicities. This behavior limits their utility as general ionization-parameter diagnostics, and we therefore do not consider them further as tracers of $Q$.

The equivalent width of O3, however, exhibits a strong dependence on the ionization parameter across subsolar metallicities, spanning nearly two orders of magnitude in EWO3 and making it a particularly sensitive diagnostic of the ionization parameter in low-metallicity systems. At $Z \geq 0.02$, the EWO3–$Q$ relation departs from simple monotonic behavior at high $Q$, reflecting changes in the ionization balance in metal-rich gas. This behavior mirrors the non-monotonic dependence of the O32 ratio on ionization parameter at high metallicity seen in classical photoionization models \citep[e.g.,][]{Kewley02}.

Among the equivalent widths considered, EWO3 exhibits the strongest and most nearly monotonic dependence on ionization parameter across the subsolar metallicity regime. Together with its observational advantages over O32, particularly in high-redshift galaxies where $\OII$ is often weak or falls outside the observable wavelength range, this behavior makes EWO3 the most promising equivalent-width ionization-parameter diagnostic for star-forming galaxies. In the remainder of this work, we therefore focus on calibrating EWO3 against the ionization parameter.

%\begin{figure}
%    \centering
%    \includegraphics{logQ.png}
%    \caption{Caption}
%    \label{fig:O32_Q}
%\end{figure}

\begin{figure}
    \centering
    \includegraphics[width=0.9\linewidth]{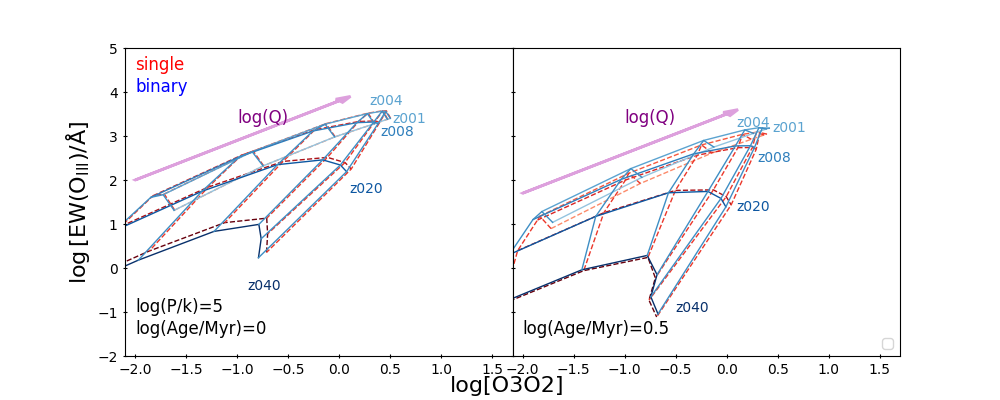}
    \caption{Equivalent width of $\mathrm{O_{III}}$ (EWO3) versus O32 ratio from our model. The left panel shows the correlation for the model assuming $\log(\mathrm{Age/Myr}) = 0$ and the right panel shows the model with $\log(\mathrm{Age/Myr}) = 0.5$. The red solid lines show the values from single stellar model while the blue dashed lines are those from the model with binary. The arrow in the left panel shows the increment in the $\log(Q)$ in the grid, which covers seven values of $\log(Q)$ ranging from 6.5 to 9.0 dex with 0.5 dex increments. The grid in the left panel is labelled with gas phase metallicity with value $Z = 0.001, 0.004, 0.008, 0.02$ and $0.04$.}
    \label{fig:EW_O32_young}
\end{figure}

\begin{figure}
    \centering
    \includegraphics[width=0.5\linewidth]{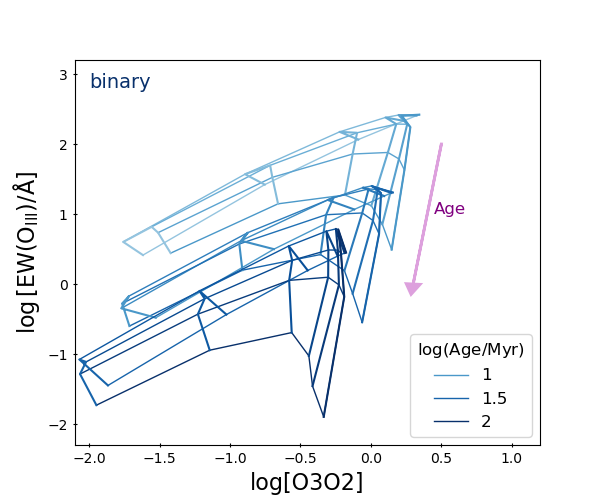}
    \caption{Equivalent width of $\mathrm{O_{III}}$ (EWO3) versus O32 ratio from our binary-inclusive model. Different color shows variation in stellar population ages, as indicated in the legend. The grids cover the $\log(Q)$ value ranging from 6.5 to 9.0 dex with 0.5 dex increment and gas phase metallicity with value $Z = 0.001, 0.004, 0.008, 0.02$ and $0.04$. }
    \label{fig:EW_O32_old}
\end{figure}

\begin{figure}
    \centering
    \includegraphics[width=0.5\linewidth]{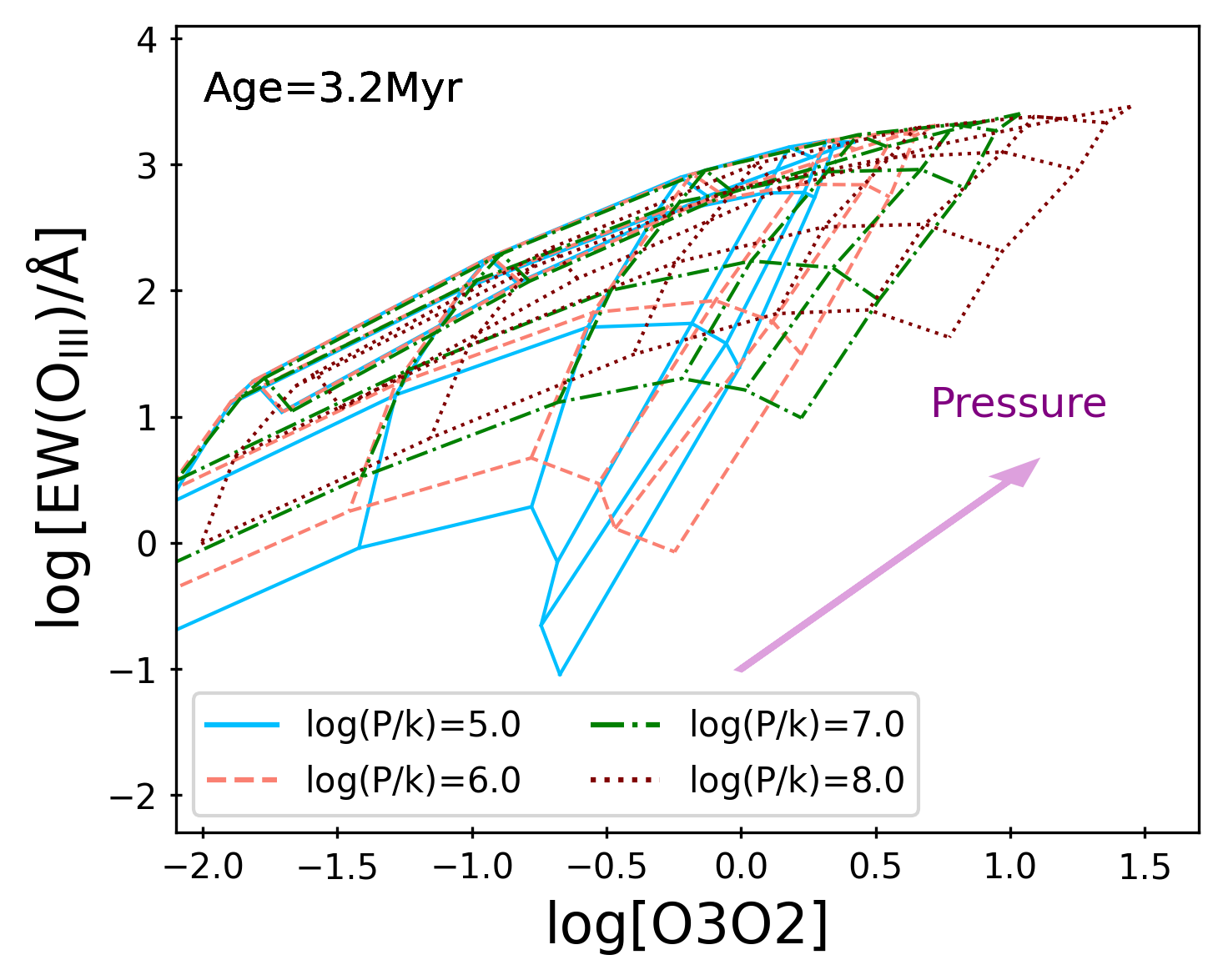}
    \caption{Equivalent width of $\OIII\lambda5007$ (EWO3) versus the O32 ratio ($\OIII\lambda5007/\OII\lambda3727$) for the binary-inclusive models at a fixed stellar population age of 3.2 Myr. Different line styles and colors indicate gas pressure, as labeled in the legend. The model grids span a range of ionization parameters (increasing from left to right) and gas-phase metallicities (increasing from top to bottom). The arrow indicates the direction of increasing gas pressure in the high-ionization-parameter regime, where increasing pressure produces simultaneous increases in both EWO3 and O32.}
    \label{fig:EW_O32_pres}
\end{figure}

\section{EWO3--O32 Diagnostic Space}
\label{sec:diagnostic}
In this section, we explore the EWO3--O32 diagnostic plane, which combines two directly observable quantities that trace the ionization state of the nebular gas. While O32 is a well-established ionization-parameter diagnostic, EWO3 is a diagnostic we propose and test in this work. Examining their joint behavior in the EWO3--O32 plane provides a natural framework for assessing the sensitivity of EWO3 relative to O32 under variations in stellar population age, metallicity, and gas pressure. We first present model predictions to establish how different physical parameters map onto this diagnostic space, and then compare these predictions with observations across a wide range of redshifts. This approach enables a direct and uniform comparison between models and data, and provides the foundation for the EWO3 calibration presented in the following section.

\subsection{Model Predictions}
\label{ssec:model_predictions}

Using the photoionization model grid described in Table~\ref{tab:model_grid}, we examine how variations in stellar population age, gas metallicity, ionization parameter, and gas pressure shape the EWO3--O32 plane.

Figures~\ref{fig:EW_O32_young}--\ref{fig:EW_O32_pres} show the distribution of model predictions in the EWO3--O32 plane. Increasing the ionization parameter primarily moves models toward the upper-right, corresponding to simultaneous increases in both EWO3 and O32, as indicated by the arrow in the both panels of Figure~\ref{fig:EW_O32_young}. In Figure~\ref{fig:EW_O32_young}, the model grids are constructed by varying $\log(Q)$ and gas metallicity while the gas pressure is held fixed at $\log(P/k)=5$. In the following subsections, we isolate the effects of stellar population age, gas metallicity, and gas pressure to determine how each parameter shapes the EWO3--O32 plane and to identify the regimes in which EWO3 provides a robust ionization-parameter diagnostic.

\subsubsection{Sensitivity to Stellar Population Age}
\label{sssec:age_sensitivity}

Figure~\ref{fig:EW_O32_young} shows the distribution of model predictions in the EWO3--O32 plane at young stellar population ages ($\log(\mathrm{Age/Myr}) = 0$ and $0.5$) for both single-star and binary-inclusive models. At these early times, the ionizing radiation field is dominated by massive O stars, and both model variants produce similar EWO3 and O32 values, resulting in only minor differences between the single-star and binary-inclusive sequences.

As the stellar population ages, the decline in ionizing photon production leads to systematic shifts in the EWO3--O32 plane. This behavior is illustrated in Figure~\ref{fig:EW_O32_old}, where binary-inclusive models continue to populate the diagnostic space while single-star models rapidly cease to produce significant nebular emission and therefore no longer yield predictions in this plane. The evolution with age causes models to move toward the bottom-left region of the EWO3--O32 plane, corresponding to simultaneous decreases in both EWO3 and O32, as indicated by the arrow in Figure~\ref{fig:EW_O32_pres}.

The sensitivity to stellar population age differs between the two diagnostics. Over the age range $\log(\mathrm{Age/Myr}) = 0$--2, EWO3 decreases by approximately 2--3 dex, whereas O32 varies by only $\sim0.5$--1 dex. This indicates that EWO3 is substantially more sensitive to stellar population age than O32, reflecting its direct dependence on the normalization of the ionizing radiation field relative to the stellar continuum.

At older stellar population ages ($\log(\mathrm{Age/Myr}) \gtrsim 1$), the dependence of EWO3 on metallicity becomes weaker. This is evident when comparing model grids at different ages: at $\log(\mathrm{Age/Myr}) = 1$ and above, the vertical spread in EWO3 across metallicity is reduced relative to that at younger ages (e.g. $\log(\mathrm{Age/Myr}) = 0.5$). In this regime, the hardness of the ionizing radiation field is increasingly governed by binary evolution processes rather than by metal opacities in stellar atmospheres. Consequently, variations in stellar population age dominate the behavior of EWO3, while metallicity plays a secondary role.

\subsubsection{Sensitivity to Gas Metallicity}
\label{sssec:metallicity_sensitivity}

Gas metallicity influences the structure of the EWO3--O32 plane by regulating both the efficiency of nebular cooling and the hardness of the ionizing radiation field due to metal opacities in the stellar atmospheres. At fixed stellar population age, gas pressure, and ionization parameter, changes in metallicity do not move models along the direction of increasing $\log(Q)$, but instead shift their position within the EWO3--O32 plane (see Figure~\ref{fig:EW_O32_young}).

At young stellar population ages, increasing metallicity initially leads to a weak rise in EWO3, which peaks at $Z \simeq 0.004$ before declining toward higher metallicities. This behavior has also been found in the relation between $[\mathrm{O\,\textsc{iii}}]$ line luminosity and metallicity predicted using photoionization models coupled with \textsc{BPASS} \citep{Xiao18}. The initial increase in EWO3 reflects enhanced cooling via collisionally excited optical transitions as the metal abundance rises. At higher metallicities, however, the dominant cooling channels shift to infrared fine-structure transitions \citep{Spitzer78}, reducing the sensitivity of optical emission lines to further increases in metal abundance. In this regime, the softer ionizing radiation field at high metallicity is increasingly unable to excite highly ionized species, leading to a decline in EWO3 \citep{Xiao18, Eldridge17}.

As metallicity increases, the O32 ratio decreases steadily. Singly ionized oxygen ($\mathrm{O^+}$), which has a lower ionization potential than doubly ionized oxygen ($\mathrm{O^{++}}$), can be produced efficiently by the softer ionizing spectra characteristic of higher-metallicity stellar populations. In contrast, the production of $\mathrm{O^{++}}$ requires harder ionizing photons, which become increasingly scarce as metallicity rises due to enhanced line blanketing and increased stellar atmospheric opacity. As a result, $[\mathrm{O\,\textsc{iii}}]$ emission weakens relative to $[\mathrm{O\,\textsc{ii}}]$ emission with increasing metallicity, driving a monotonic decline in the O32 ratio.

Overall, both EWO3 and O32 decrease with increasing metallicity, but with different sensitivities. Across the full metallicity range explored ($Z = 0.001$--0.04), metallicity variations span $\sim2$--4~dex in EWO3, while O32 varies by only $\sim1$~dex. However, the majority of the EWO3 variation occurs at near-solar and super-solar metallicities; when considering only the subsolar regime ($Z \leq 0.008$), the metallicity-driven spread in EWO3 is reduced to $\lesssim0.5$~dex, comparable to that of O32. In this regime, metallicity acts primarily as a secondary shaping parameter in the EWO3--O32 plane, preserving the diagnostic utility of EWO3 as a tracer of the ionization parameter.

\subsubsection{Sensitivity to Gas Pressure}
\label{sssec:pressure_sensitivity}

Figure~\ref{fig:EW_O32_pres} shows the distribution of model predictions in the EWO3–O32 plane for different gas pressures at a fixed stellar population age of 3.2 Myr, where ionization parameter increases from left to right and metallicity increases from top to bottom. The effect of gas pressure on the model grids depends on both ionization parameter and metallicity. At low ionization parameter (left side of the grid), increasing gas pressure produces little change at low metallicity, but introduces a vertical shift toward higher EWO3 at higher metallicity, with minimal change in O32. In contrast, at high ionization parameter (right side of the grid)—particularly at moderate to high metallicity—increasing pressure produces simultaneous increases in both EWO3 and O32, resulting in a diagonal shift toward higher values of both diagnostics, as indicated by the arrow in the figure.

At high ionization parameter and low metallicity, however, variations in gas pressure produce larger changes in O32 than in EWO3, indicating that the line ratio remains more responsive to pressure in this regime while the equivalent width varies only weakly. This behavior demonstrates that the diagnostic sensitivity of EWO3 and O32 depends on the underlying physical conditions, with each tracer responding differently across the parameter space and exhibiting complementary sensitivity in different regimes. As discussed previously, EWO3 shows a weaker dependence on secondary parameters in the low-metallicity regime, reinforcing its robustness as a diagnostic in these conditions.

This behavior reflects the role of gas density and pressure in regulating nebular cooling. In metal-rich gas, cooling is dominated by infrared fine-structure transitions, which suppress optical emission lines such as O3. Increasing gas pressure enhances collisional excitation and reduces the relative importance of infrared cooling, allowing optical O3 emission to become more prominent. As a result, high-pressure conditions can partially offset the suppression of O3 emission in metal-rich environments, while pressure-induced changes remain more modest at low metallicity and low ionization parameter, where optical cooling channels are already efficient.

Overall, gas pressure introduces systematic but non-uniform shifts in the EWO3--O32 plane. While both diagnostics become more sensitive to pressure at high ionization parameter, O32 exhibits a substantially stronger response than EWO3 in the low-metallicity, high-ionization regime. Consequently, EWO3 provides a more robust ionization-parameter diagnostic than O32 under the low-metallicity, high-ionization conditions commonly found in high-redshift galaxies, owing to its weaker dependence on gas pressure. These results motivate the pressure-dependent calibration of EWO3 presented in the following section.

\begin{figure}
    \centering
    \includegraphics[width=1.0\linewidth]{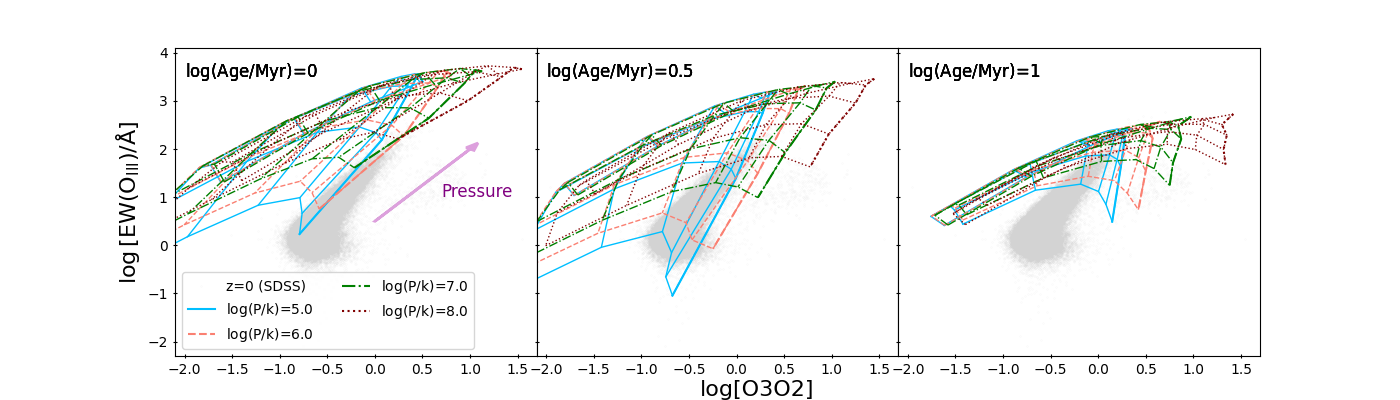}
    \caption{EWO3–O32 plane for photoionization models compared with SDSS star-forming galaxies (grey points). Panels show different stellar population ages: $\log(\mathrm{Age/Myr}) = 0$ (left), 0.5 (middle), and 1 (right). Model grids span a range of ionization parameters and gas-phase metallicities, with different colors and line styles indicating gas pressure ($\log(P/k)=5$–8). Models with $\log(\mathrm{Age/Myr}) = 0.5$ provide the best agreement with the SDSS distribution.}
    \label{fig:vs_local}
\end{figure}

\begin{figure} 
    \centering
    \includegraphics[width=0.8\linewidth]{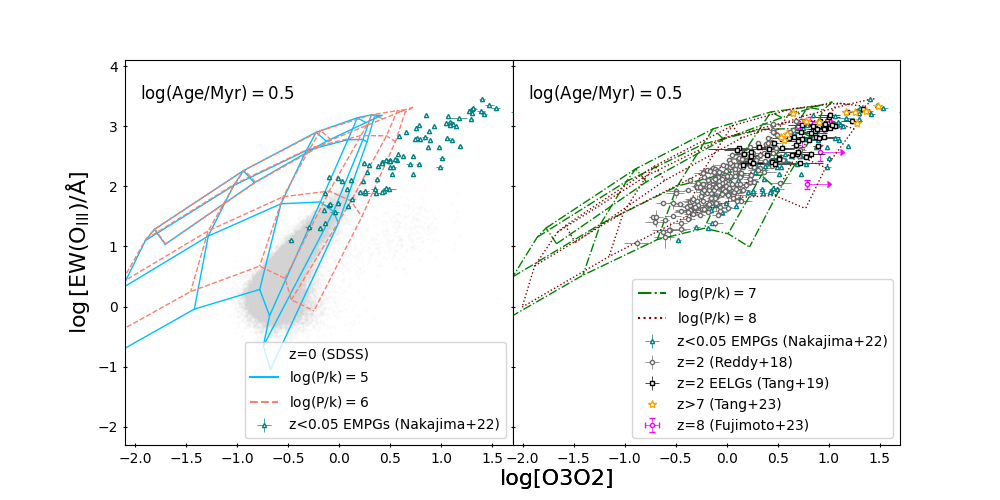}
    \caption{Distribution of observed galaxies in the EWO3–O32 plane compared with photoionization model predictions at a fixed stellar age of 3.2 Myr. Left panel: low-pressure models with $\log(P/k)=5$ and 6 overlaid with local galaxies (SDSS and EMPGs). Right panel: high-pressure models with $\log(P/k)=7$ and 8 overlaid with intermediate- and high-redshift galaxies. In both panels, model grids span a range of metallicities and ionization parameters. Local galaxies are well reproduced by low-pressure models, while high-redshift galaxies occupy regions of the diagnostic plane that require substantially higher gas pressures. Open symbols with error bars denote observational measurements from \citet{Nakajima22, Reddy18, Tang19, Fujimoto23, Tang23}, as indicated in the legend.}
    \label{fig:EW_O32_calibration}
\end{figure}

\subsection{Comparison with Observations}
\label{ssec:comparison}

We now compare the model predictions in the EWO3--O32 plane with observations of star-forming galaxies across a wide range of redshifts. Guided by the parameter sensitivities established in Section~\ref{ssec:model_predictions}, we use the EWO3--O32 diagnostic space to identify the physical conditions required to reproduce different galaxy populations. Figure~\ref{fig:EW_O32_calibration} presents a direct comparison between photoionization model grids and observed galaxies, enabling a unified interpretation of local and high-redshift systems within the same framework.

\subsubsection{Local Galaxies}
\label{sssec:local}

Figure~\ref{fig:vs_local} shows the comparison between photoionization model predictions and SDSS star-forming galaxies for different stellar population ages and gas pressures. Models with $\log(\mathrm{Age/Myr}) = 1$ clearly do not match the overall SDSS distribution and are therefore not considered as fiducial models.

Models with a very young stellar population age of $\log(\mathrm{Age/Myr}) = 0$ systematically overpredict EWO3, particularly at high metallicity, and fail to reproduce the low-EWO3 end of the SDSS distribution. Increasing the gas pressure does not resolve this discrepancy, as higher pressure shifts the model grids toward larger EWO3 values, further exacerbating the mismatch. This indicates that the ionizing radiation fields in typical local galaxies are softer than those produced by an extremely young ($\sim$1~Myr) stellar population.

In contrast, models with $\log(\mathrm{Age/Myr})=0.5$ provide the best agreement with the SDSS galaxies. At this age, the observed distribution is well reproduced by the lower-pressure models, $\log(P/k)=5$ and 6, with galaxies occupying moderate metallicities ($Z\gtrsim0.004$) and ionization parameters in the range $7.5\lesssim\log Q\lesssim9$, consistent with typical conditions in nearby $\HII$ regions \citep[e.g.,][]{Pacifici12, Xiao18}. This age is representative of actively star-forming regions and provides the best overall match to the SDSS distribution among the stellar population models explored in this work.

The comparison suggests that local star-forming galaxies are well described by young ($\sim3.2$ Myr) stellar populations and relatively low gas pressures. In Section \ref{sec:calibration}, we quantify this comparison using the EWO3 calibration to determine the pressure that best reproduces the O32-derived ionization parameters.

\subsubsection{Non-Local and Extreme Galaxies}
\label{sssec:nonlocal}

Figure~\ref{fig:EW_O32_calibration} compares photoionization model predictions with observed galaxies that occupy the high-EWO3, high-O32 region of the diagnostic plane, including extremely metal-poor galaxies \citep[EMPGs,][]{Nakajima22}, intermediate-redshift star-forming galaxies, and high-redshift systems. These populations are grouped together based on their location in the EWO3--O32 plane, as they lie outside the locus of typical SDSS galaxies and trace more extreme nebular excitation conditions.

For the high-redshift CEERS galaxies \citep{Tang23}, only the blended EW($[\mathrm{O\,\textsc{iii}}]+\mathrm{H}\beta$) is available due to aperture-loss uncertainties in the \textit{JWST}/NIRSpec micro-shutter array, which prevent reliable measurement of individual line equivalent widths. However, the $[\mathrm{O\,\textsc{iii}}]/\mathrm{H}\beta$ flux ratio is well determined. To place these galaxies on the EWO3--O32 plane, we estimate EW($[\mathrm{O\,\textsc{iii}}]\,\lambda5007$) by partitioning the blended equivalent width using the observed $[\mathrm{O\,\textsc{iii}}]/\mathrm{H}\beta$ ratio and adopting the canonical 1:3 flux ratio between $[\mathrm{O\,\textsc{iii}}]\,\lambda4959$ and $\lambda5007$. Assuming a uniform continuum level across the blend, the contribution of $\lambda5007$ is given by
\[
\mathrm{EW}_{5007} =
\frac{3R}{3R + R + 1}\,
\mathrm{EW}_{[\mathrm{O\,III}]+\mathrm{H}\beta},
\]
where $R = F_{[\mathrm{O\,III}]5007}/F_{\mathrm{H}\beta}$. Although approximate, this reconstruction is sufficient for placing the CEERS galaxies in the EWO3--O32 diagnostic diagrams for comparison with other high-redshift samples.

EMPGs overlap with intermediate- and high-redshift galaxies in diagnostic space, exhibiting elevated O32 values relative to the bulk of the SDSS population. In the low-pressure model grid (left panel of Figure~\ref{fig:EW_O32_calibration}), EMPGs lie largely outside the model predictions because of their elevated O32 values, while their EWO3 values remain broadly consistent with the model range. In the high-pressure model grid (right panel of Figure~\ref{fig:EW_O32_calibration}), their EWO3 and O32 values overlap the region occupied by the high-redshift samples, demonstrating that extreme excitation is not unique to the early Universe. In the case of EMPGs, this behaviour is likely linked to their very low metallicities, which produce intrinsically hard ionizing radiation fields capable of generating large EWO3 values.

Intermediate- and high-redshift galaxies, including the MOSDEF sample \citep{Reddy18}, extreme emission-line galaxies \citep{Tang19}, and the highest-redshift CEERS galaxies \citep{Tang23, Fujimoto23}, occupy the upper-right region of the EWO3--O32 plane, reaching the highest observed values of both diagnostics. This region is characterized by simultaneously large EWO3 and high O32 ratios, and is clearly offset from the locus of typical local SDSS galaxies.

High-pressure photoionization models provide a direct explanation for the location of these galaxies in diagnostic space. Increasing the gas pressure shifts model grids toward higher EWO3 and O32, allowing the observed loci of MOSDEF, EELG, and CEERS galaxies to be reproduced. In contrast, low-pressure models cannot reach the observed combination of EWO3 and O32 values, regardless of the assumed ionization parameter. Even in high-pressure models, young stellar populations ($\log(\mathrm{Age/Myr}) \le 0.5$) remain required; models with older stellar populations fail to reproduce the observed excitation levels at intermediate and high redshift. Among the model grids explored, only those combining elevated gas pressure with moderately young stellar populations reproduce the full range of observed EWO3 and O32 values.

Taken together, these results demonstrate that high EWO3 values can arise through multiple physical pathways. Extremely metal-poor galaxies achieve large EWO3 at relatively low gas pressure due to intrinsically hard ionizing radiation fields, while high-redshift galaxies reach similarly extreme EWO3 values through elevated gas pressure at more moderate metallicities. This highlights that gas pressure governs the most extreme nebular conditions traced by EWO3 and O32, with lower metallicity systems requiring less pressure to reach a given level of excitation than more metal-rich systems.

\section{Calibration of EWO3} \label{sec:calibration}

The strong correlation between EWO3 and O32 identified in Section~\ref{sec:EW}
suggests that EWO3 may be used as an ionization-parameter diagnostic, particularly
at intermediate and high redshift where O32 cannot be measured. In this section, we 
construct model-based calibrations of EWO3 as a function of the ionization parameter 
and test these calibrations against observations from $z \sim 0$ to $z \sim 8$.

\subsection{Calibration Grids and Surface Fits}
\label{sec:calibration_fits}

\begin{figure}
    \centering
    \includegraphics[width=0.8\linewidth]{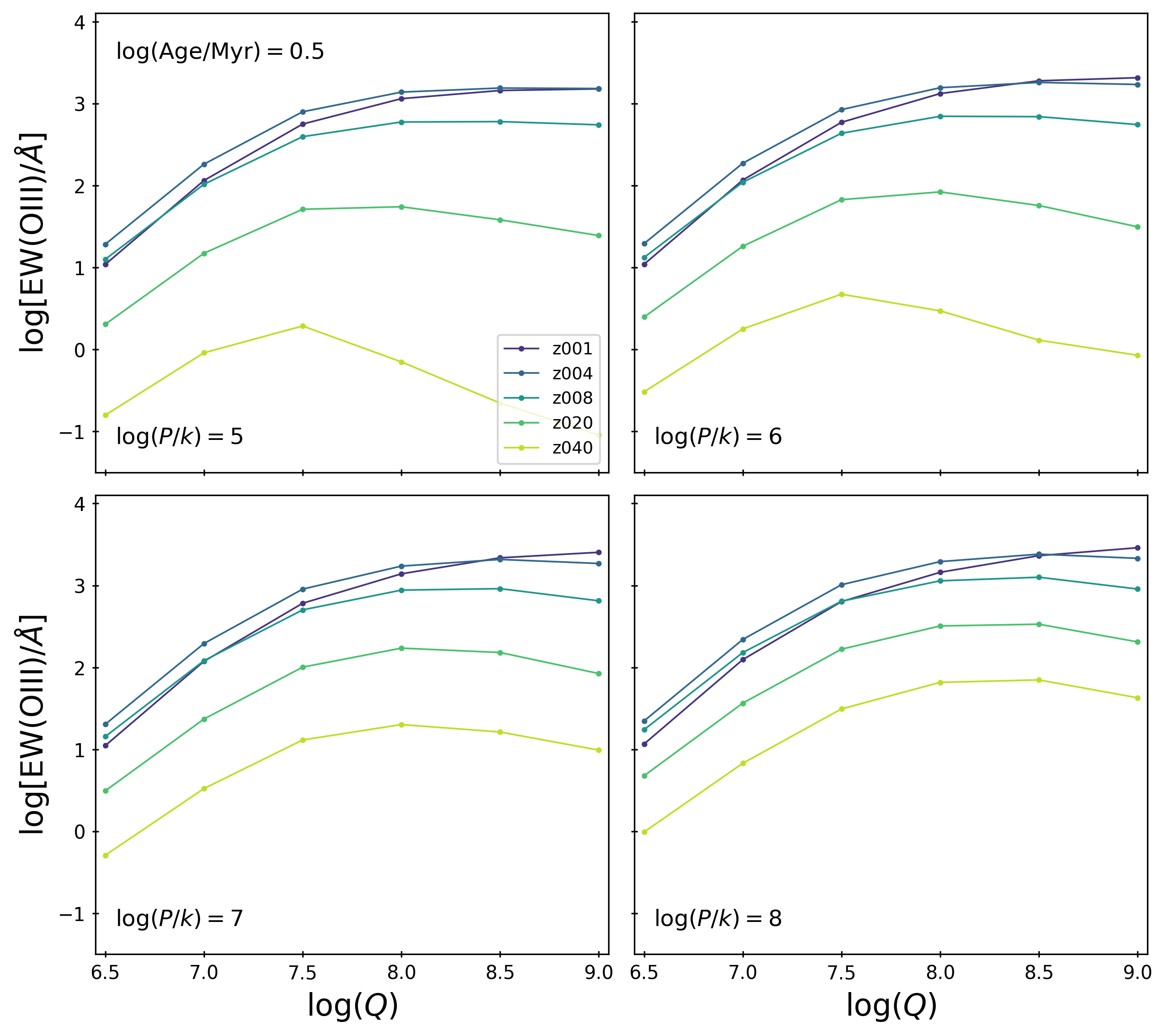}

    \caption{
Dependence of EWO3 on ionization parameter for different gas pressures at a fixed stellar population age of $\log(\mathrm{Age/Myr}) = 0.5$. The four panels show photoionization model grids with $\log(P/k)=5$, 6, 7, and 8. Curves correspond to gas-phase metallicities $Z=0.001$, 0.004, 0.008, 0.02, and 0.04, equivalent to $12+\log(\mathrm{O/H})=7.57$, 8.17, 8.47, 8.87, and 9.17, respectively.}
    \label{fig:EW_Q_calibration}
\end{figure}

To construct the EWO3 ionization-parameter diagnostic, we adopt the photoionization model grids presented in Section~\ref{ssec:model_predictions}. Figure~\ref{fig:EW_Q_calibration} shows that the behavior of the EWO3--$\log(Q)$ relation depends on both gas pressure and metallicity. At near-solar and super-solar metallicities, the lower-pressure grids ($\log(P/k)=5$ and 6) become increasingly non-monotonic, producing double-valued solutions that complicate the inversion from EWO3 to ionization parameter. In contrast, the higher-pressure grids ($\log(P/k)=7$ and 8) remain single-valued over the full ionization-parameter range explored. 

We therefore constructed separate surface-fit calibrations for each pressure grid (log(P/k)=5, 6, 7, and 8), spanning the full metallicity range of the models and ionization parameters up to log(Q)=9.0. The resulting calibrations were subsequently validated against observed samples spanning the local Universe to $z\sim8$, including local star-forming galaxies, extremely metal-poor galaxies, and high-redshift galaxies, to determine which pressure grid most accurately reproduces ionization parameters derived from the O32 diagnostic across different physical environments.

\medskip
\noindent
\textbf{Surface-fit calibration.}
We construct separate bivariate surface-fit calibrations for each pressure grid,
$\log(P/k)=5$, 6, 7, and 8. Because the EWO3--$\log(Q)$ relation becomes
non-monotonic for the $\log(P/k)=5$ and 6 grids, separate surface fits are
derived for the lower- and upper-ionization branches. In contrast, the
$\log(P/k)=7$ and 8 grids remain single-valued over the calibration domain and
are each described by a single surface fit.

For each calibration, we parameterize the relationship between EWO3,
metallicity, and ionization parameter using the bivariate polynomial
\begin{equation}
    \log(Q) = a_0 + a_1 \log(\mathrm{EWO3})
                + a_2 Z
                + a_3 \log(\mathrm{EWO3})^2
                + a_4 Z^2
                + a_5 \log(\mathrm{EWO3})\,Z,
    \label{eq:EW_surface}
\end{equation}
where $Z \equiv 12+\log(\mathrm{O/H})$. The best-fitting coefficients for each
pressure grid are listed in Table~\ref{tab:EW_surface_coeffs}. These surface
fits are used throughout the remainder of the paper to compute
$\log(Q)_{\mathrm{EWO3}}$.

\medskip
\noindent
\textbf{Branch selection using O3/H$\beta$.}
The $\log(P/k)=5$ and 6 calibrations exhibit a turnover in the EWO3--$\log(Q)$ relation near $\log(Q)\approx7.5$, requiring galaxies to be
assigned to either the lower- or upper-ionization branch before evaluating the
surface fit. For this purpose, we use the
$\log(\mathrm{O3}/\Hb)$ ratio, which generally increases with ionization
parameter \citep[e.g.,][]{Kewley19}. Within our model grid,
$\log(\mathrm{O3}/\Hb)$ provides a clear separation between the lower- and
upper-ionization branches of the EWO3--$\log(Q)$ relation.
branches of the EWO3--$\log(Q)$ relation.

Using the photoionization models, we derive the branch-selection boundary
\begin{equation}
    \log(\mathrm{O3/H}\beta)_{\rm crit} = -0.691\,Z + 5.671 ,
\end{equation}
where $Z \equiv 12+\log(\mathrm{O/H})$. Galaxies lying above (below) this
boundary are assigned to the upper (lower) branch of the $\log(P/k)=5$ and 6
calibrations. The $\log(P/k)=7$ and 8 model grids remain single-valued over the full calibration range and therefore require no branch selection.

\medskip
\noindent
\textbf{Choice of pressure grid.}
The appropriate pressure grid may be selected using electron-density
measurements when available, or guided by the characteristic physical
environment of the galaxy. Based on the observational validation presented in
Section~\ref{sec:calibration_test}, we find that the
$\log(P/k)=6$ calibration provides the best agreement with O32-derived
ionization parameters for local star-forming galaxies and extremely metal-poor
galaxies, while the $\log(P/k)=7$ calibration is preferred for the
high-redshift galaxy population.

\begin{table*}
\centering

\caption{Coefficients for the bivariate surface-fit calibration
$\log(Q)=a_0 + a_1\log(\mathrm{EWO3}) + a_2 Z + a_3\log(\mathrm{EWO3})^2
+ a_4 Z^2 + a_5 \log(\mathrm{EWO3})Z$, where
$Z \equiv 12+\log(\mathrm{O/H})$. Separate fits are provided for each
pressure grid. The $\log(P/k)=5$ and 6 grids are represented by separate
lower- and upper-ionization branches, while the $\log(P/k)=7$ and 8 grids
are described by single surface fits.}
\label{tab:EW_surface_coeffs}

\begin{tabular}{lcccccc}
\hline
\hline
Regime & $a_0$ & $a_1$ & $a_2$ & $a_3$ & $a_4$ & $a_5$ \\
\hline
$\log(P/k)=5$ (low-$Q$)   &
87.416968 &
$-0.187402$ &
$-20.246321$ &
0.018607 &
1.252424 &
0.093712 \\

$\log(P/k)=5$ (upper-$Q$) &
158.941110 &
$-8.992706$ &
$-35.479087$ &
0.443182 &
2.075839 &
0.960946 \\

$\log(P/k)=6$ (low-$Q$) &
79.284582 &
$-0.436017$ &
$-18.145423$ &
0.020976 &
1.117634 &
0.121858 \\

$\log(P/k)=6$ (upper-$Q$) &
5.584319 &
$-0.124345$ &
$-0.096340$ &
0.272512 &
0.046371 &
$-0.068140$ \\

$\log(P/k)=7$ &
129.726450 &
$-3.279830$ &
$-30.159565$ &
0.173858 &
1.828536 &
0.444751 \\

$\log(P/k)=8$ &
111.478269 &
$-3.962197$ &
$-25.298818$ &
0.254734 &
1.512361 &
0.480587 \\
\hline
\end{tabular}
\end{table*}

\subsection{Testing the EWO3 Ionization-Parameter Diagnostic}
\label{sec:calibration_test}

We now evaluate the EWO3 ionization-parameter diagnostic by applying each
pressure-dependent calibration to observed galaxy samples spanning the local
Universe to $z\sim8$. The resulting ionization parameters,
$\log(Q)_{\rm EWO3}$, are compared with values derived from the traditional O32 diagnostic, $\log(Q)_{\rm O32}$, using the iterative calibration from \citet{Kobulnicky04}. Gas-phase metallicities are likewise determined from the \citet{Kobulnicky04} R23 calibration to ensure a consistent comparison between the two diagnostics.

We compare the $\log(P/k)=5$, 6, 7, and 8 calibrations for three galaxy
populations representing different physical environments: local SDSS
star-forming galaxies, extremely metal-poor galaxies (EMPGs) from
\citet{Nakajima22}, and intermediate- to high-redshift galaxies from
\citet{Reddy18}, \citet{Tang23}, and \citet{Fujimoto23}. The preferred pressure grid for each population is identified by minimizing the systematic offset and scatter relative to $\log(Q)_{\rm O32}$.

\begin{figure*}
\centering
\includegraphics[width=\textwidth]{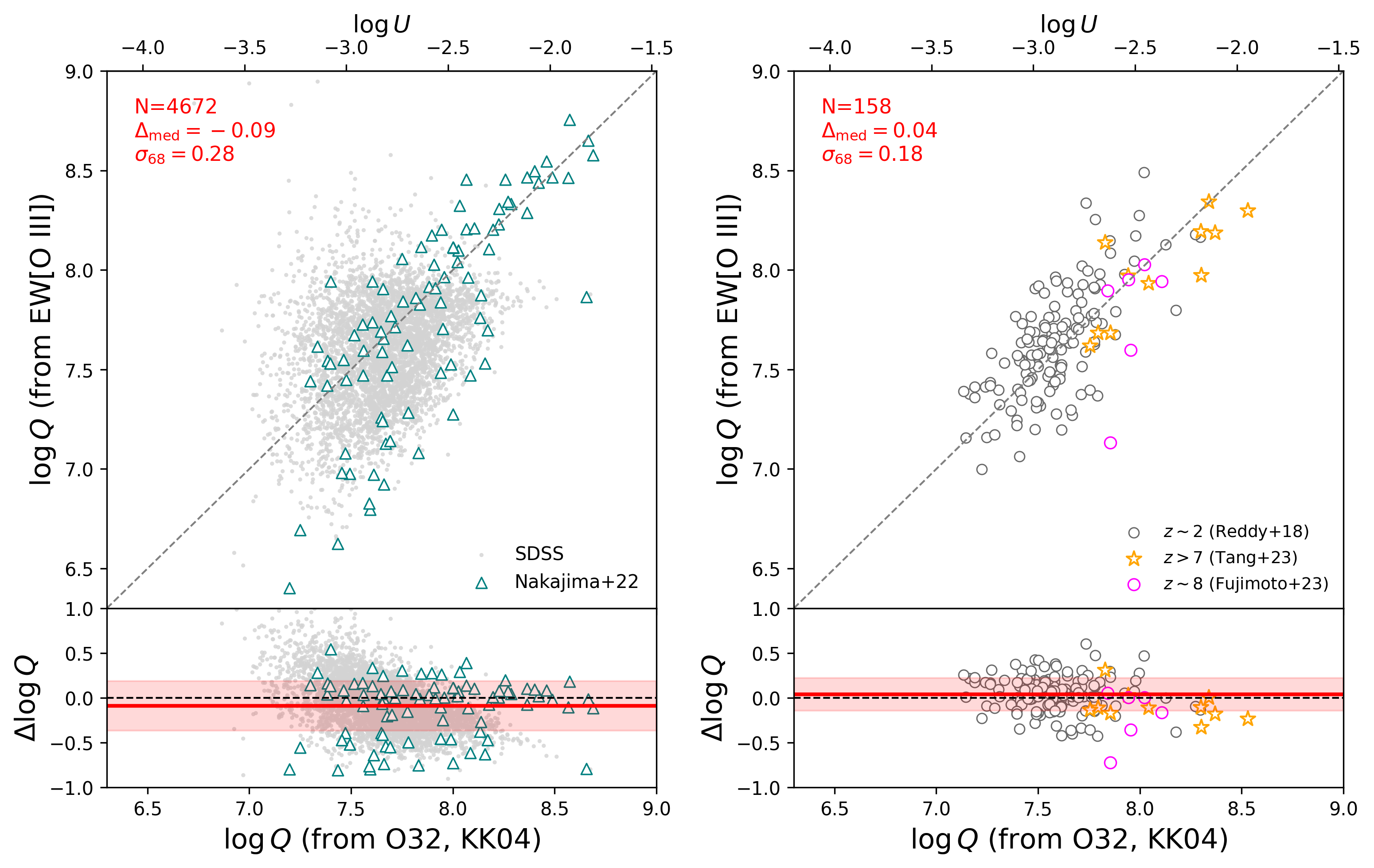}
\caption{
Validation of the EWO3 ionization-parameter diagnostic against ionization parameters derived from the O32 ratio using the iterative KK04 calibration. The upper panels compare $\log(Q)_{\mathrm{EWO3}}$ with $\log(Q)_{\mathrm{O32}}$, while the lower panels show the residuals,
$\Delta\log(Q)=\log(Q)_{\mathrm{EWO3}}-\log(Q)_{\mathrm{O32}}$. The dashed lines indicate the one-to-one relation (upper panels) and zero residual (lower panels). \textit{Left:} Local star-forming galaxies from SDSS (grey points) and extremely metal-poor galaxies from \citet{Nakajima22} (cyan triangles), evaluated using the $\log(P/k)=6$ calibration. The combined sample shows excellent agreement with the O32 diagnostic, with a median offset of $-0.09$ dex and $\sigma_{68}=0.28$ dex. \textit{Right:} Intermediate- and high-redshift galaxies from \citet{Reddy18}, \citet{Tang23}, and \citet{Fujimoto23}, evaluated using the $\log(P/k)=7$ calibration. This calibration provides the best agreement for high-redshift galaxies, yielding a median offset of $0.04$ dex and $\sigma_{68}=0.18$ dex.
}
\label{fig:EW_O32_comparison}
\end{figure*}

Figure~\ref{fig:EW_O32_comparison} presents the preferred EWO3 pressure-dependent calibrations identified from the comparison with O32-derived ionization parameters. For local star-forming galaxies and the extremely metal-poor galaxies (EMPGs) of \citet{Nakajima22}, the $\log(P/k)=6$ calibration provides the best agreement with $\log(Q)_{\mathrm{O32}}$, yielding a combined median offset of $-0.09$ dex and a scatter of $\sigma_{68}=0.28$ dex. The close agreement over a wide range of ionization parameters demonstrates that intermediate-pressure photoionization models provide an appropriate description of the ionized gas conditions in these nearby galaxy populations.

For the intermediate- and high-redshift galaxies from \citet{Reddy18}, \citet{Tang23}, and \citet{Fujimoto23}, the $\log(P/k)=7$ calibration provides the closest agreement with the O32 diagnostic. The resulting median offset is only $0.04$ dex with a scatter of $\sigma_{68}=0.18$ dex, indicating that a moderately higher-pressure model better reproduces the nebular conditions characteristic of high-redshift star-forming galaxies.

The preferred pressure grids therefore differ between galaxy populations. While the $\log(P/k)=6$ calibration is most suitable for local star-forming galaxies and EMPGs, the $\log(P/k)=7$ calibration is preferred for intermediate- and high-redshift galaxies. This progression is consistent with the expectation that the characteristic pressure of H\,\textsc{ii} regions increases toward the more compact, intensely star-forming systems observed at high redshift.

At high redshift, where $\OII\lambda3727$ is frequently inaccessible and O32 cannot be measured directly, the EWO3 diagnostic provides a practical alternative for estimating the ionization parameter from rest-frame optical spectra. Moreover, EWO3 is less sensitive than O32 to variations in gas pressure under the low-metallicity, high-ionization conditions commonly found in high-redshift galaxies, making it a good alternative ionization-parameter diagnostic in this regime. The excellent agreement with O32-derived ionization parameters demonstrated here shows that EWO3 can reliably recover the ionization parameter when an appropriate pressure calibration is adopted. We recommend the $\log(P/k)=6$ calibration for local star-forming galaxies and the $\log(P/k)=7$ calibration for intermediate- and high-redshift galaxies.

%\label{sec:EWO3-age}
%\begin{figure}
%    \centering
%    \includegraphics{images/EWO3_O32_age_b.png}
%    \caption{Caption}
%    \label{fig:EWO3_O32}
%\end{figure}

%\begin{figure}
%    \centering
%    \includegraphics{images/N2Ha_Z.png}
%    \caption{Caption}
%    \label{fig:N2Ha_Z}
%\end{figure}

%\begin{figure}
%    \centering
%    \includegraphics{images/N2O3_Z.png}
%    \caption{Caption}
%    \label{fig:N2O3_Z}
%end{figure}

%\begin{figure}
%    \centering
%    \includegraphics{images/O3Hb_Z.png}
%    \caption{Caption}
%    \label{fig:O3Hb_Z}
%\end{figure}

\section{Conclusion} \label{sec:summary}

We have presented predictions for the equivalent width of $[\mathrm{O\,\textsc{iii}}]\,\lambda5007$ (EWO3) by coupling the photoionization model \textsc{MAPPINGS} with the stellar population synthesis model \textsc{BPASS}. By varying stellar population and nebular parameters, we investigated how these factors influence the resulting emission-line properties. Our main conclusions are as follows:

\begin{itemize}
    \item We explore two variants of the \textsc{BPASS} model: (i) single-star instantaneous burst models and (ii) binary-inclusive instantaneous burst models. The ionizing radiation fields from both models are similar for young, metal-poor populations where massive O stars dominate the production of ionizing photons. In the single-star models, the ionizing spectra decline sharply after $\sim4$--$5$~Myr due to the death of massive stars. In contrast, binary interactions in the binary-inclusive models allow stellar rejuvenation, enabling the population to continue emitting ionizing radiation to later ages. As a result, the ionizing spectra from binary-inclusive models are also less sensitive to metallicity than those from single-star models, as the effects of binary evolution compete with metal-dependent stellar winds in governing stellar evolution (see Figure~\ref{fig:bpass_Z}).

    \item We propose EWO3 as an alternative ionization-parameter diagnostic, particularly for intermediate- and high-redshift observations where the spectral coverage is limited and the commonly used O32 ratio cannot be measured. Using photoionization models, we compare the behavior of EWO3 and O32 over a range of stellar population and nebular parameters. Both diagnostics increase with ionization parameter and show similar dependencies on metallicity, stellar population age, and gas pressure. However, EWO3 is less sensitive than O32 to variations in gas pressure under the low-metallicity, high-ionization conditions commonly found in high-redshift galaxies, making it a more robust ionization-parameter diagnostic in this regime.

    \item Comparing our models with observations of local SDSS galaxies and extremely metal-poor galaxies (EMPGs), we find that model grids adopting a stellar population age of $\log(\mathrm{Age/Myr})=0.5$ and a gas pressure of $\log(P/k)=6$ provide the best agreement with ionization parameters derived from the O32 diagnostic. This result suggests that intermediate-pressure photoionization models provide a more representative description of the ionized gas conditions in nearby star-forming galaxies than either lower- or higher-pressure grids.

    \item Intermediate- and high-redshift galaxies exhibit systematically higher EWO3 and O32 values than local galaxies, reflecting the more extreme physical conditions within their H\,\textsc{ii} regions. Among the pressure-dependent calibrations considered, the $\log(P/k)=7$ models provide the closest agreement with O32-derived ionization parameters, indicating that moderately higher-pressure photoionization models are required to reproduce the nebular conditions characteristic of high-redshift star-forming galaxies.
    
    \item We present pressure-dependent calibrations of EWO3 as an ionization-parameter diagnostic based on photoionization models spanning $\log(P/k)=5$--8. Validation against observed galaxies demonstrates that the $\log(P/k)=6$ calibration is preferred for local star-forming galaxies and EMPGs, while the $\log(P/k)=7$ calibration provides the best agreement for intermediate- and high-redshift galaxies. These calibrations enable reliable estimates of the ionization parameter from EWO3 when the commonly used O32 diagnostic cannot be measured, such as when $[\mathrm{O\,\textsc{ii}}]$ lies outside the observed wavelength range.

\end{itemize}

%\begin{figure}
%    \centering
%    \includegraphics[width=0.5\linewidth]{images/EWO3_O32_POSTER.png}
%    \caption{EWO3 versus O32 ratio from our model adopting BPASS binary stellar population with $\log(\mathrm{Age/Myr}) = 0.5$. Cyan grid shows the model with a pressure of $\log \mathrm{(P/k) =5}$  while maroon grid shows that with $\log \mathrm{(P/k) =8}$. Arrows mark the effect of varying different parameters as labelled. The open symbols with errorbar are the observational values from \cite{Reddy18}, \cite{Tang19} and \cite{Fujimoto23}, as indicated in the legend.}
%    \label{fig:conclusion}
%\end{figure}
%% IMPORTANT! The old "\acknowledgment" command has be depreciated. It was
%% not robust enough to handle our new dual anonymous review requirements and
%% thus been replaced with the acknowledgment environment. If you try to 
%% compile with \acknowledgment you will get an error print to the screen
%% and in the compiled pdf.
%% 
%% Also note that the akcnowlodgment environment does not support long amounts of text. If you have a lot of people and institutions to acknowledge, do not use this command. Instead, create a new \section{Acknowledgments}.

\begin{acknowledgments}
DPT thanks Yifei Jin, Peixin Zhu, and Phoebe Heretz for helpful discussions on emission-line diagnostics and comparisons between photoionization models and observations.

Funding for SDSS-III has been provided by the Alfred P. Sloan Foundation, the Participating Institutions, the National Science Foundation, and the U.S. Department of Energy Office of Science. The SDSS-III website is \url{http://www.sdss3.org/}.
\end{acknowledgments}

%% To help institutions obtain information on the effectiveness of their 
%% telescopes the AAS Journals has created a group of keywords for telescope 
%% facilities.
%
%% Following the acknowledgments section, use the following syntax and the
%% \facility{} or \facilities{} macros to list the keywords of facilities used 
%% in the research for the paper.  Each keyword is check against the master 
%% list during copy editing.  Individual instruments can be provided in 
%% parentheses, after the keyword, but they are not verified.

\vspace{5mm}

%% Similar to \facility{}, there is the optional \software command to allow 
%% authors a place to specify which programs were used during the creation of 
%% the manuscript. Authors should list each code and include either a
%% citation or url to the code inside ()s when available.

\software{\textsc{MAPPINGS V} \citep{Sutherland18}, \textsc{BPASS v2.2} \citep{Stanway18, Eldridge17}}

%% Appendix material should be preceded with a single \appendix command.
%% There should be a \section command for each appendix. Mark appendix
%% subsections with the same markup you use in the main body of the paper.

%% Each Appendix (indicated with \section) will be lettered A, B, C, etc.
%% The equation counter will reset when it encounters the \appendix
%% command and will number appendix equations (A1), (A2), etc. The
%% Figure and Table counter will not reset.

%% For this sample we use BibTeX plus aasjournals.bst to generate the
%% the bibliography. The sample631.bib file was populated from ADS. To
%% get the citations to show in the compiled file do the following:
%%
%% pdflatex sample631.tex
%% bibtext sample631
%% pdflatex sample631.tex
%% pdflatex sample631.tex

\bibliography{bibliography}{}
\bibliographystyle{aasjournal}

%% This command is needed to show the entire author+affiliation list when
%% the collaboration and author truncation commands are used.  It has to
%% go at the end of the manuscript.
%\allauthors

%% Include this line if you are using the \added, \replaced, \deleted
%% commands to see a summary list of all changes at the end of the article.
%\listofchanges

\end{document}